\documentclass[11pt]{article}
\usepackage{amssymb,amsmath,amsthm,amscd,latexsym}
\usepackage{mathrsfs}
\usepackage{mathrsfs}
\usepackage{amsfonts}
\usepackage{amsmath}
\usepackage{amssymb}
\usepackage{amscd}
 \usepackage{color}
 \usepackage{bm}
\usepackage{extpfeil}
\usepackage{url}
\usepackage{hyperref}

\renewcommand{\paragraph}{\roman{paragraph}}
\input cyracc.def

\newcommand{\F}{\mathbb{F}}

\newtheorem{thm}{\bfseries  Theorem}[section]
\newtheorem{lem}[thm]{\bfseries   Lemma}
\newtheorem{coro}[thm]{\bfseries   Corollary}

\newtheorem{ex}[thm]{\bfseries   Example}
\newtheorem{rem}[thm]{\bfseries   Remark}
\newtheorem{Conjecture}[thm]{\bfseries   Conjecture}

\begin{document}
%\begin{CJK*}{GBK}{song}\CJKtilde
\title{\bf A new method for constructing linear codes with small hulls
\thanks{This research is supported by the National Natural Science Foundation of China
under Grant 11771007 and Grant 61572027.}
}
\author{Liqin Qian\thanks{Liqin Qian, Department of Mathematics, Nanjing University of Aeronautics and Astronautics, Nanjing, Jiangsu, 210007, China, {\tt qianliqin\_1108@163.com}},
Xiwang Cao\thanks{Xiwang Cao, Department of Mathematics, Nanjing University of Aeronautics and Astronautics, Nanjing, Jiangsu, 210007, China; Laboratory of Information Security,
Institute of Information Engineering, Chinese Academy of Sciences, Beijing 100042, China, {\tt xwcao@nuaa.edu.cn}}, Wei Lu\thanks{School of Mathematics, Southeast University,
Nanjing, Jiangsu, 211189, China, {\tt  luwei1010@seu.edu.cn}}, Patrick Sol\'e\thanks{I2M(CNRS, Aix-Marseille University, Centrale Marseille), Marseilles, France, {\tt sole@enst.fr}} %, Sihem Mesnager\thanks{Sihem Mesnager, Department of Mathematics, University of Paris VIII, 93526 Saint-Denis, France; University of Paris XIII, CNRS, LAGA UMR 7539, Sorbonne Paris Cit\'e, 93430 Villetaneuse,
%France; Telecom ParisTech, 75013 Paris, France, {\tt sihem.mesnager@telecom-paris.fr}}
}

\date{}
\maketitle
%\end{CJK*}
%\begin{CJK}{GBK}{song}
\begin{abstract}
The hull of a linear code over finite fields is the intersection of the code and its dual, which was introduced by Assmus and Key.
%Hulls of linear codes have been of interest and extensively studied in recent years due to their rich algebraic structures and wide applications.
In this paper, we develop a method to construct  linear codes with trivial hull ( LCD codes) and one-dimensional hull by employing the positive characteristic analogues of Gauss sums.
These codes are quasi-abelian, and sometimes doubly circulant.
Some sufficient conditions for a linear code to be an LCD code (resp. a linear code with one-dimensional hull) are presented.
It is worth mentioning that we present a lower bound on the minimum distances of the constructed linear codes. As an application, using these conditions,
we obtain some optimal or almost optimal LCD codes (resp. linear codes with one-dimensional hull) with respect to the online Database of Grassl.
%The method of constructions of this paper is presented by employing the homomorphism from finite fields into finite fields.
%There are two major ingredients into this paper. The first ingredient is to generalizes the results of Li and Zeng when $(p,r)=1$.
%It turns out that our constructions are more general and direct than previous work on small hulls of linear codes.
\end{abstract}
{\bf Keywords:} Linear codes,  hull of a code, LCD codes, Gauss sums, quasi-abelian codes, double circulant codes\\
{\bf MSC(2010):} 94B05, 11T24, 11T71

\section{Introduction}

The  hull of a linear code $C$ over a finite field is defined to be $${\rm Hull}(C):=C\cap C^\perp.$$ It is clear that Hull$(C)$ is also linear.
It is easy to see that a linear code $C$ is  self-orthogonal if and only if the dimension of Hull$(C)$ is  the dimension of $C$, i.e., Hull$(C)=C,$
and it is  Linear Complementary Dual (LCD) if and only if the dimension of Hull$(C)$ is zero, i.e., Hull$(C)=\{{\bm 0}\}.$
Specifically, a linear code $C$ is  self-dual if and only if the dimension of Hull$(C)$ is $\frac{n}{2}$ for even $n$, where $n$ is the length of $C$.

Hulls of linear codes have been introduced to classify finite projective planes in \cite{AK}. Later, it turned out that hulls of linear codes play a vital role in determining the complexity of some algorithms for checking permutation equivalence of two linear codes and computing the automorphism group of a linear code \cite{L1, L2, S1,S2}. It has been shown that these algorithms are always effective when the dimension of the hull is small. Due to their wide applications, some families of linear codes with special hulls such as LCD codes and linear codes with one-dimensional hull have been of interest and extensively studied \cite{CLM,LL,LL1,ST,LZ,SH,LS}. It is worth noting that the equivalence of many types of codes with LCD codes has been extensively studied.  Jin and Xing \cite{J1} showed that an algebraic geometry code over $\mathbb{F}_{2^m}(m \geq7)$ is equivalent to an LCD code. Moreover, a celebrated result was presented in \cite{CMTQ}, which proved that any linear code over $\mathbb{F}_q~(q >3)$ is equivalent to an LCD code. These codes are practically useful in communications systems, various applications, and link with other objects as shown in \cite{CG, CMTQ,CMTQ1,CMTQ2,J} and references therein. Consequently, it is of interest to study hulls, families of linear codes with small hulls. What needs to be emphasized is that Li and Zeng \cite{LZ} constructed linear codes with one-dimensional hull by utilizing quadratic Gaussian sums from quadratic number fields and Carlet, Li and Mesnager \cite{CLM} constructed LCD codes and linear codes with one-dimensional hull by employing character sums in semi-primitive case from cyclotomic fields and multiplicative subgroups of finite fields. They have made a lot of contributions in this regard.

Inspired by the above research work, we construct LCD codes and codes with one-dimensional hull dimension, by using an analogue of Gauss sums where both the corresponding additive and multiplicative character take their
values in a finite field instead of the complex numbers. This method generalizes previous work \cite{LZ, QC1,QC2}. Moreover, we consider the order $N\geq 2$ of the homomorphism, while
\cite{LZ} only considers $N=2$. It turns out that our constructions are more general and direct than previous work on small hulls of linear codes. It is worth observing that
we obtain some optimal or almost optimal LCD codes and linear codes with one-dimensional hull from our constructions. Compared with \cite{LZ}, the linear codes we constructed may be
new when $N> 2$ in the sense. Furthermore, we also present a lower bound on the minimum distances of the codes presented in this paper. These codes
have a lot of built-in symmetry: they are quasi-abelian of index $2$  in general \cite{JL}, and
double circulant in many cases.
% that the generator matrix is different.

The rest of this paper is organized as follows. Section 2 gives the preliminaries. In Section 3, we give two concrete homomorphisms from a
finite field into a finite field and present the idea of constructing linear codes determined by a special generator matrix. In Sections 4 and 5, we investigate LCD codes
and linear codes with one-dimensional hull by employing these two homomorphisms from a finite field to a finite field, respectively. In addition, we present some examples
of optimal or almost optimal LCD codes and linear codes with one-dimensional hull. In Section 6, we present a lower bound on the minimum distances of the constructed linear codes.
%use an inductive method to determine the minimum distances of the constructed linear codes.
Section 7 concludes the article.
\section{Preliminaries}
In this section, we introduce some notation and results in order for the exposition in this paper to be self-contained, which will be useful later.
\subsection{Codes}
Let $q$ be a power of a prime $p$ and $\F_q$ denote the finite field with $q$ elements.
For a positive integer $n$, a linear code of length $n$ over $\F_q$ is defined to be a subspace of the $\F_q$-vector spaces $\F_q^n.$ A linear code $C$ of length $n$ over $\F_q$ is called an $[n,k,d]_q$ code if its $\F_q$-dimension is $k$ and the minimum Hamming distance of $C$ is $d$. If $C$ is an $[n,k,d]$ code, then from the Singleton bound, its minimum distance is bounded above by $d\leq n-k+1$. A code meeting the above bound is called Maximum Distance Separable (MDS). A code is called almost MDS if its minimum distance is one less than the MDS case. For ${\bm u}:=(u_1,u_2,\cdots, u_n)$ and ${\bm v}:=(v_1,v_2,\cdots, v_n)$ in $\F_q^n$, the  inner product of ${\bm u}$ and ${\bm v}$ is defined to be $\langle{\bm u},{\bm v}\rangle:=\sum\limits_{i=1}^nu_iv_i.$ The  dual $C^\perp$ of a linear code $C$ of length $n$ over $\F_q$ is defined to be the set $C^\perp=\{{\bm v}\in \F_q^n|\langle{\bm c},{\bm v}\rangle=0~{\rm for~all~}{\bm c}\in C\}.$
A linear code $C$ is said to be  self-orthogonal if $C\subseteq C^\perp$ and it is said to be  self-dual if $C=C^\perp$. A linear code $C$ is said to be  linear complementary dual (LCD) code if $C\cap C^\perp=\{0\}.$

\subsection{Homomorphisms}
Starting from this subsection till the end of this paper, we let $\F_{r^m}$ denote the finite field of order $r^m$, where $r$ is a prime number and $m$ is a positive integer. Let $\F_{r^m}^*=\F_{r^m}\backslash \{0\}.$ Let
%$\F_{q}$ be the finite field with $q$ elements, where $q$ is a power of a prime $p$ and let
$\overline{\F}_q$ be the algebraic closure of the finite field $\F_q.$

Let $\varphi$ be a homomorphism from $\F_{r^m}^*$ into $\overline{\F}_q^*$, that is, a mapping from $\F_{r^m}^*$ into $\overline{\F}_q^*$ with $\varphi(xy)=\varphi(x)\varphi(y)$ for all $x,y\in \F_{r^m}^*.$ Define $\overline{\varphi}(x):=\varphi(x^{-1}).$ Let $\varphi_0$ be the trivial homomorphism, which is defined by $\varphi_0(x)=1$ for all $x\in \mathbb{F}_{r^m}^*.$

The following lemma gives the orthogonality relations of the homomorphism $\varphi$.
\begin{lem}\label{clem1} Let $\varphi$ be defined as above. Then we have
\begin{equation*}
\sum\limits_{x\in \mathbb{F}_{r^m}^*}\varphi(x)=\begin{cases}
\emph{ }r^m-1,  ~{\rm if}~\varphi=\varphi_0;\\
   \emph{ }0, ~~~~~~~~{\rm if}~\varphi\neq\varphi_0.\\
\end{cases}
\end{equation*}
\end{lem}
\begin{proof} The proof is similar to that of \cite[Theorem 5.4]{LNC} and omitted here.
\end{proof}
Let $\chi$ be a homomorphism from $\F_{r^m}$ into $\overline{\F}_q^*$, that is, a mapping from $\F_{r^m}$ into $\overline{\F}_q^*$ with $\chi(x+y)=\chi(x)\chi(y)$ for all $x,y\in \F_{r^m}.$ Define $\overline{\chi}(x):=\chi(-x).$ Let $\chi_0$ be the trivial homomorphism, which is defined by $\chi_0(x)=1$ for all $x\in \mathbb{F}_{r^m}.$

We also have the following lemma, which presents the orthogonality relations of the homomorphism $\chi$.
\begin{lem}\label{lem1} Let $\chi$ be defined as above. Then we have
\begin{equation*}
\sum\limits_{x\in \mathbb{F}_{r^m}}\chi(x)=\begin{cases}
\emph{ }r^m,  ~{\rm if}~\chi=\chi_0;\\
   \emph{ }0, ~~~{\rm if}~\chi\neq\chi_0.\\
\end{cases}
\end{equation*}
\end{lem}
\begin{proof} The proof is similar to that of \cite[Theorem 5.4]{LNC} and omitted here.
\end{proof}
\subsection{Some results for the sum $g(\varphi,\chi)$}
Let $\varphi$ and $\chi$ be defined as Subsection 2.1. Then we define the sums $$g(\varphi, \chi)=\sum\limits_{x\in \mathbb{F}_{r^m}^*}\varphi(x)\chi(x)$$ and $$\overline{g(\varphi, \chi)}=g(\overline{\varphi},\overline{\chi})=\sum\limits_{x\in \mathbb{F}_{r^m}^*}\varphi(x^{-1})\chi(-x).$$
The following results show the value of the sum $g(\varphi,\chi)$.
\begin{lem}\label{lemp}
Let $\varphi$ and $\chi$ be defined as Subsection 2.1. Then the sum $g(\varphi, \chi)$ satisfies
\begin{equation*}
g(\varphi, \chi)=\begin{cases}
\emph{ }r^m-1,  ~~&{\rm if}~\varphi=\varphi_0~and~\chi=\chi_0;\\
   \emph{ }~-1, ~~~~&{\rm if}~\varphi=\varphi_0~and~\chi\neq\chi_0; \\
   \emph{ }~~~~0,~~~~~&{\rm if}~\varphi\neq\varphi_0~and~\chi=\chi_0.\\
\end{cases}
\end{equation*}
\end{lem}
\begin{proof} The conclusion follows directly from Lemmas \ref{clem1} and \ref{lem1}.
\end{proof}
\begin{lem}\label{lem2} Let $\varphi$ and $\chi$ be defined as Subsection 2.1. If $\varphi\neq\varphi_0~and~ \chi\neq\chi_0$, then $$g(\varphi, \chi)\overline{g(\varphi, \chi)}=r^m\in \mathbb{F}_p.$$
\end{lem}
\begin{proof} For $\varphi\neq\varphi_0~{\rm and}~\chi\neq\chi_0$, we get \begin{eqnarray*}
              % \nonumber to remove numbering (before each equation)
                g(\varphi, \chi)\overline{g(\varphi, \chi)} &=& \sum\limits_{x\in  \mathbb{F}_{r^m}^*}\varphi(x)\chi(x)\sum\limits_{y\in  \mathbb{F}_{r^m}^*}\varphi(y^{-1})\chi(-y) \\
                 &=& \sum\limits_{x,y\in  \mathbb{F}_{r^m}^*}\varphi(xy^{-1})\chi(x-y) \\
                &\xlongequal{x\longrightarrow xy}& \sum\limits_{x,y\in  \mathbb{F}_{r^m}^*}\varphi(x)\chi(y(x-1)) \\
                &=& \sum\limits_{x\in  \mathbb{F}_{r^m}^*}\varphi(x)\sum\limits_{y\in  \mathbb{F}_{r^m}^*}\chi(y(x-1))\\
                &=& \varphi(1)\sum\limits_{y\in  \mathbb{F}_{r^m}^*}\chi(0)+\sum\limits_{x\in  \mathbb{F}_{r^m}^*\backslash \{1\}}\varphi(x)\sum\limits_{y\in  \mathbb{F}_{r^m}^*}\chi(y(x-1))\\
                &=& r^m-1-\sum\limits_{x\in  \mathbb{F}_{r^m}^*\backslash \{1\}}\varphi(x)\\
                &=&r^m-\sum\limits_{x\in \mathbb{F}_{r^m}^*}\varphi(x)\\
                &=& r^m.
              \end{eqnarray*}This completes the proof of this lemma.
\end{proof}
The study of the behavior of the sum $g(\varphi,\chi)$ under various transformations of the $\varphi$ or $\chi$ leads to a number of useful identities.
\begin{lem}\label{lem3} Let $\varphi$ and $\chi$ be defined as Subsection 2.1. Then we have the following results.
\begin{enumerate}
 % \item [(1)] $g(\varphi_j, \chi_{ab})=\varphi_j(b^{-1})g(\varphi_j, \chi_a)$ for $a\in \mathbb{F}_{r^m}, b\in \mathbb{F}_{r^m}^*$;
  \item [(1)] $g(\varphi, \overline{\chi})=\varphi(-1)g(\varphi, \chi)$;
  \item [(2)] $g(\overline{\varphi}, \chi)=\varphi(-1)\overline{g(\varphi, \chi)}$;
  \item [(3)] $g(\varphi, \chi)g(\overline{\varphi}, \chi)=\varphi(-1)r^m$ for $\varphi\neq \varphi_0$~and~$\chi\neq\chi_0$;
  %\item [(4)] $g(\varphi^r, \chi_a)=g(\varphi_j, \chi_{\sigma(a)})$ for $a\in \mathbb{F}_{r^m}$ and $\sigma(a)=a^r$;
  \item [(4)] $(g(\varphi,\chi))^{p^s}=g(\varphi^{p^s},\chi^{p^s})$, where $p$ is the characteristic of $\F_q$ and $s$ is a positive integer.
  %Moreover, if $N|(p^s+1)$
\end{enumerate}
\end{lem}
\begin{proof} The results of (1)-(3) are obvious by the definition $g(\varphi, \chi)$ and Lemma \ref{lem2}. Next, we prove the result of (4). Combined with the definitions of $\varphi$ and $\chi$, we have

$(g(\varphi,\chi))^{p^s}=\left(\sum\limits_{x\in \mathbb{F}_{r^m}^*}\varphi(x)\chi(x)\right)^{p^s}=\sum\limits_{x\in \mathbb{F}_{r^m}^*}(\varphi(x))^{p^s}(\chi(x))^{p^s}=\sum\limits_{x\in \mathbb{F}_{r^m}^*}\varphi^{p^s}(x)\chi^{p^s}(x)=g(\varphi^{p^s},\chi^{p^s}).$
\end{proof}
\begin{rem} The $\varphi$ and $\chi$ in Section 2.1 are not the usual
%not what we normally call
multiplicative and additive characters, respectively. Moreover, the $g(\varphi,\chi)$ is also not the usual Gauss sums. However, we can prove that the sum $g(\varphi,\chi)$ has similar properties to Gauss sums (see Lemmas \ref{lemp} and \ref{lem3}(1-3)). The definition of the sum $g(\varphi,\chi)$ may have been studied before, but we haven't found any relevant references.
\end{rem}
\subsection{On characterizations of LCD codes and codes having one-dimensional hull}
In this paper, we consider the constructions of linear codes with small hull, mainly refer to LCD codes and linear codes with one-dimensional hull. We will characterize when a linear code is an LCD code or a linear code with one-dimensional hull. We next present two lemmas for this purpose.

A complete characterization of LCD codes via the nonsingularity of their generator matrices was employed in \cite{CG,M}, which provides a sufficient and necessary condition for a linear code to be an LCD code.
\begin{lem}\label{lemm1} \cite{CG,M}
Let $C$ be an $[n,k]$ linear code over $\F_q$ with generator matrix $G=[I_k, {P}]$. Then the code $C$ is LCD if and only if $I_k+{P}{P}^T$ is nonsingular, i.e., $-1$ is not an eigenvalue of the matrix ${P}{P}^T$, where ${P}^T$ denotes the transpose of ${P}.$
\end{lem}
We also have the following lemma on a linear code having one-dimensional hull, which provides an idea to construct linear codes with one-dimensional hull by using the eigenvalues of the generator matrices.
\begin{lem}\label{lem4}\cite{CLM,LZ} Let $C$ be an $[n,k]$ linear code over $\F_q$ with generator matrix $G=[I_k, {P}]$. Then the code $C$ has one-dimensional hull if the matrix ${P}{P}^T$ has an eigenvalue $-1$ with $($algebraic$)$ multiplicity $1$.
\end{lem}

\section{Linear codes associated with homomorphisms}
In this section, we construct the linear codes by using the two homomorphisms in Section 2.1.

Let $r$ be a prime number and $m$ a positive integer. $\mathbb{F}_{r^m}$ denotes the finite field of order $r^m$. Let $\mathbb{F}_{r^m}^*=\mathbb{F}_{r^m}\backslash\{0\}$ and $\mathbb{F}_{r^m}^*=\langle \alpha\rangle$, where $\alpha$ is a fixed primitive element of $\mathbb{F}_{r^m}^*$. Assume that $N>1$ is a positive integer and $N|(r^m-1)$. Let $q$ be a power of $p$, where $p$ is a prime number. Assume that $N|(q-1)$.
%and $4|(q-1)$.
Let $\mathbb{F}_q^*=\langle \beta\rangle$, where $\beta$ is a fixed primitive element of $\mathbb{F}_q^*.$ For the sake of convenience, we let $u=\beta^{\frac{q-1}{N}}$.
% and $v=\beta^\frac{q-1}{4}$.
Define the function
\begin{eqnarray}\label{eq1}
% \nonumber to remove numbering (before each equation)
  && \varphi:\mathbb{F}_{r^m}^*\longrightarrow {\mathbb{F}}_q^*, \varphi(\alpha^k)=u^{k},
\end{eqnarray}
where $0\leq k\leq r^m-2.$
%It follows from the definition, the order of $\varphi_j$ is $N$, i.e., $\varphi_j^N=\varphi_0$. The function $\varphi_j$ is a homomorphism from $\mathbb{F}_{r^m}^*$ to $\mathbb{F}_q^*$.
It is easy to know that $\varphi$ is a homomorphism of order $N$. Define the kernel of the homomorphism $\varphi$ is the set $${\rm ker}(\varphi):=\{\alpha^k, 0\leq k\leq r^m-2: \varphi(\alpha^k)=1\}=\langle \alpha^N\rangle.$$
%$\varphi_j(\alpha^{k_1}\alpha^{k_2})=\varphi_j(\alpha^{k_1})\varphi_j(\alpha^{k_2})$ for $0\leq k_1, k_2\leq r^m-2$.

Assume that $(p,r)=1$. Then there exists a positive integer $t$ such that $r|(q^t-1)$. Let $\mathbb{F}_{q^t}^*=\langle \gamma\rangle$ and $\zeta=\gamma^{\frac{q^t-1}{r}}$, where $\gamma$ is a fixed primitive element of $\mathbb{F}_{q^t}^*$. For any $a\in \mathbb{F}_{r^m},$ we define
\begin{eqnarray}\label{eq2}
% \nonumber to remove numbering (before each equation)
  && \chi_a: \mathbb{F}_{r^m}\longrightarrow \overline{\mathbb{F}}_{q}^*, \chi_a(x)=\zeta^{{\rm Tr}_r^{r^m}(ax)}, x\in \mathbb{F}_{r^m},
\end{eqnarray}
where ${\rm Tr}_r^{r^m}$ denotes the trace function from $\mathbb{F}_{r^m}$ onto $\mathbb{F}_r$. It is easy to know that $\chi_a$ is a homomorphism. %$\chi_a(x_1+x_2)=\chi_a(x_1)\chi_a(x_2)$ for $x_1,x_2\in \mathbb{F}_{r^m}.$
%For convenience, we define $\chi:=\chi_1$.
It follows from the definition of $\chi_a$ that
\begin{eqnarray}\label{eqn8}
% \nonumber to remove numbering (before each equation)
 g(\varphi, \chi_{ab}) &=& \overline{\varphi}(b)g(\varphi, \chi_a)
\end{eqnarray} for $a\in\F_{r^m}$ and $b\in \F_{r^m}^*.$

Fix $v\in \F_q.$ Let $\mathbb{F}_{r^m}=\{x_i: 1\leq i\leq r^m\}$. Define the $r^m\times r^m$ matrix $P=(p_{ij})\in M_{r^m}(\mathbb{F}_q)$ by setting $p_{ij}=\rho(x_j-x_i)$, where
\begin{equation}\label{mat}
\rho(x_j-x_i)=\begin{cases}
\emph{ }\varphi(x_j-x_i),  ~{\rm if}~i\neq j;\\
   \emph{ }v, ~~~~~~~~~~~~~{\rm if}~i=j.\\
\end{cases}
\end{equation}
For any $a\in \F_{r^m}$, set $\eta_a:=(\chi_a(x_1), \chi_a(x_2), \cdots, \chi_a(x_{r^m}))^T$, where $``T"$ denotes the transpose operator. Then the $i$th component of $P\eta_a$ is
$$\sum\limits_{j=1}^{r^m}\rho(x_j-x_i)\chi_a(x_j)=\sum\limits_{x\in \mathbb{F}_{r^m}}\rho(x-x_i)\chi_a(x)\xlongequal{y:=x-x_i}\sum\limits_{y\in \mathbb{F}_{r^m}}\rho(y)\chi_a(y+x_i)= \sum\limits_{y\in  \mathbb{F}_{r^m}}\rho(y)\chi_a(y)\chi_a(x_i).$$
Hence, $P\eta_a=\left(\sum\limits_{y\in  \mathbb{F}_{r^m}}\rho(y)\chi_a(y)\right)\eta_a$ and $\eta_a$ is an eigenvector of $P$.

Similarly, the $i$th component of $P^T\eta_a$ is $$\sum\limits_{j=1}^{r^m}\rho(x_i-x_j)\chi_a(x_j)=\sum\limits_{x\in \mathbb{F}_{r^m}}\rho(x_i-x)\chi(x)\xlongequal{y:=x_i-x}\sum\limits_{y\in \mathbb{F}_{r^m}}\rho(y)\chi_a(x_i-y)=\sum\limits_{y\in \mathbb{F}_{r^m}}\rho(y){\chi}_a(-y)\chi_a(x_i).$$
Hence, $P^T\eta_a=\left(\sum\limits_{y\in  \mathbb{F}_{r^m}}\rho(y)\chi_a(-y)\right)\eta_a$ and $\eta_a$ is also an eigenvector of $P^T$.

Next, we will prove that the $r^m$ vectors $\{\eta_a:=(\chi_a(x_1), \chi_a(x_2), \cdots, \chi_a(x_{r^m}))^T: a\in \mathbb{F}_{r^m}\}$ are linearly independent over $\overline{\mathbb{F}}_q$.

Suppose that $\sum\limits_{a\in \mathbb{F}_{r^m}}k_a \eta_a= \mathbf{0}$, where $k_a\in \overline{\mathbb{F}}_q$. Then we have
\begin{eqnarray*}
% \nonumber to remove numbering (before each equation)
  && \sum\limits_{a\in \mathbb{F}_{r^m}}k_a(\chi_a(x_1), \chi_a(x_2), \cdots, \chi_a(x_{r^m}))^T= \mathbf{0}, \\
  &\Longrightarrow& \left( \sum\limits_{a\in \mathbb{F}_{r^m}}k_a\chi_a(x_1),  \sum\limits_{a\in \mathbb{F}_{r^m}}k_a\chi_a(x_2), \cdots,  \sum\limits_{a\in \mathbb{F}_{r^m}}k_a\chi_a(x_{r^m})\right)^T=\mathbf{0}.
\end{eqnarray*}
%\begin{eqnarray*}
               % \nonumber to remove numbering (before each equation)
               %   \sum\limits_{a\in \mathbb{F}_{r^m}}k_a(\chi_a(x_1), \chi_a(x_2), \cdots, \chi_a(x_{r^m}))^T&=& \mathbf{0},~{\rm i.e.,} \\
                % ( \sum\limits_{a\in \mathbb{F}_{r^m}}k_a\chi_a(x_1),  \sum\limits_{a\in \mathbb{F}_{r^m}}k_a\chi_a(x_2), \cdots,  \sum\limits_{a\in \mathbb{F}_{r^m}}k_a\chi_a(x_{r^m}))^T&=& \mathbf{0}.
              % \end{eqnarray*}
Hence, $\sum\limits_{a\in \mathbb{F}_{r^m}}k_a\chi_a(x_i)=0$ for any $1\leq i\leq r^m$.

Given an element $a_0\in \mathbb{F}_{r^m}$, we have
\begin{eqnarray*}
% \nonumber to remove numbering (before each equation)
  &&\sum\limits_{a\in \mathbb{F}_{r^m}}k_a\chi_a(x_i)\chi_{a_0}(-x_i)=0~{\rm for~any}~1\leq i\leq r^m,\\
  &\Longrightarrow&\sum\limits_{a\in \mathbb{F}_{r^m}}k_a\chi_1((a-a_0)x_i)= 0 ~{\rm for~any}~1\leq i\leq r^m,\\
  &\Longrightarrow&\sum\limits_{x\in \mathbb{F}_{r^m}}\sum\limits_{a\in \mathbb{F}_{r^m}}k_a\chi_1((a-a_0)x)= 0, \\
  &\Longrightarrow&\sum\limits_{a\in \mathbb{F}_{r^m}}k_a\sum\limits_{x\in \mathbb{F}_{r^m}}\chi_1((a-a_0)x)=0.
\end{eqnarray*}
By Lemma \ref{lem1}, we obtain $k_{a_0}r^m=0$ and then $k_{a_0}=0$ by $(r,p)=1$. Because $a_0$ is arbitrary, we have $k_a=0$ for any $a\in \mathbb{F}_{r^m}$. Hence, the $r^m$ vectors $\{\eta_a:=(\chi_a(x_1), \chi_a(x_2), \cdots, \chi_a(x_{r^m}))^T: a\in \mathbb{F}_{r^m}\}$ are linearly independent over $\overline{\mathbb{F}}_q$.

Therefore, the multisets $\left\{\sum\limits_{y\in  \mathbb{F}_{r^m}}\rho(y)\chi_a(y): a\in \mathbb{F}_{r^m}\right\}$ and $\left\{\sum\limits_{y\in  \mathbb{F}_{r^m}}\rho(y)\chi_a(-y): a\in \mathbb{F}_{r^m}\right\}$ present all eigenvalues of the matrix $P$ and ${P}^T$, respectively.

To sum up, $PP^T\eta_a=P\left(\sum\limits_{y\in  \mathbb{F}_{r^m}}\rho(y)\chi_a(-y)\right)\eta_a=\left(\sum\limits_{y\in  \mathbb{F}_{r^m}}\rho(y)\chi_a(y)\sum\limits_{y\in  \mathbb{F}_{r^m}}\rho(y)\chi_a(-y)\right)\eta_a$. Then the multiset $\left\{\lambda_a:=\sum\limits_{y\in  \mathbb{F}_{r^m}}\rho(y)\chi_a(y)\sum\limits_{y\in  \mathbb{F}_{r^m}}\rho(y)\chi_a(-y): a\in \mathbb{F}_{r^m}\right\}$ presents all eigenvalues of the matrix $PP^T$ and $\{\eta_a: a\in \mathbb{F}_{r^m}\}$ presents all eigenvectors of $PP^T.$

 Let the symbols be the same as above. According to Lemma \ref{lem3}(1), we obtain
\begin{eqnarray*}
% \nonumber to remove numbering (before each equation)
  \lambda_a &=& \sum\limits_{y\in  \mathbb{F}_{r^m}}\rho(y)\chi_a(y)\sum\limits_{y\in  \mathbb{F}_{r^m}}\rho(y)\chi_a(-y) \\
   &=&\left(v+\sum\limits_{y\in  \mathbb{F}_{r^m}^*}\varphi(y)\chi_a(y)\right)\left(v+\sum\limits_{y\in  \mathbb{F}_{r^m}^*}\varphi(y)\chi_a(-y)\right) \\
   &=& (v+g(\varphi,\chi_a))(v+g(\varphi,\overline{\chi}_a))\\
   &=& v^2+vg(\varphi,\chi_a)+vg(\varphi,\overline{\chi}_a)+g(\varphi,\chi_a)g(\varphi,\overline{\chi}_a)\\
    &=& v^2+vg(\varphi,\chi_a)+\varphi(-1)vg(\varphi,\chi_a)+\varphi(-1)(g(\varphi,\chi_a))^2\\
    &=& v^2+(1+\varphi(-1))vg(\varphi,\chi_a)+\varphi(-1)(g(\varphi,\chi_a))^2.
\end{eqnarray*}
Hence, all eigenvalues of $PP^T$ are given by the multiset
\begin{eqnarray}\label{den4}
% \nonumber to remove numbering (before each equation)
  &&\{\lambda_a:=v^2+(1+\varphi(-1))vg(\varphi,\chi_a)+\varphi(-1)(g(\varphi,\chi_a))^2: a\in \F_{r^m} \}.
\end{eqnarray}
\hspace*{0.5cm}Let $C:=C_{(\varphi,v)}$ be a linear code over $\F_q$ with generator matrix $G=[I_{r^m}, P].$ Then $C$ is a $[2r^m, r^m]$ linear code over $\F_q$. In Section 4, we construct LCD codes according to Lemma \ref{lemm1}. In Section 5, we construct linear codes with one-dimensional hull by Lemma \ref{lem4}.
%consider adding some conditions such that $C$ is an LCD code, and in Section 5 we consider adding some conditions such that $C$ is a linear code with one-dimensional hull.
\section{The constructions of LCD codes}

In this section, we present two simple constructions of LCD codes by using the two homomorphisms (\ref{eq1}) and (\ref{eq2}). When $m=1$ these codes are double circulant. In general, they are
quasi-abelian of index $2$ as $\F_q[H]-$submodules of $\F_q[H]^2$ with $H$ the additive group of $\F_{r^m}$ \cite{JL}.\\
\emph{\textbf{Construction A.} Define $\rho(0)=v=0$. We then obtain a $r^m\times r^m$ matrix $P=(p_{ij})$ by $$p_{ij}=\rho(x_j-x_i),$$ which is defined as (\ref{mat}). It follows from (\ref{den4}) that all eigenvalues of $PP^T$ are given by
\begin{equation}\label{eqn5}
\lambda_a=\begin{cases}
\emph{ }0,  ~~~~~~~~~~~~~~~~~~~~~~{\rm if}~a=0;\\
   \emph{ }\varphi(-1)(g(\varphi,\chi_a))^2, ~~~{\rm if}~a\in \mathbb{F}_{r^m}^*. \\
\end{cases}
\end{equation}}
\hspace*{0.5cm}The following theorem gives the sufficient conditions for linear codes to be LCD codes by Construction A.
\begin{thm}\label{thm22}
Let $r$ be a prime number and $m$ be a positive integer. Assume that $N>1$ is a positive integer and $N|(r^m-1)$. Let $q$ be a power of prime $p$ and $(p,r)=1$. Assume that $N|(q-1)$.
%he $r^m\times r^m$ matrix $P=(p_{ij})\in M_{r^m}(\mathbb{F}_q)$ is defined as (\ref{mat}).
Let $C:=C_{(\varphi,0)}$ be the linear code over $\mathbb{F}_q$ with generator matrix $[I_{r^m}, P]$. Then we have the following.
\begin{enumerate}
  \item [(1)] If there exists a positive integer $s$ such that $N|(p^s-1)$ and $\varphi(p^{2s})\neq 1$, then $C$ is a $[2r^m, r^m]$ LCD code over $\mathbb{F}_q$. In particular, if $\varphi(q^{2})\neq 1$, then $C$ is an $[2r^m, r^m]$ LCD code over $\mathbb{F}_q$.
  \item [(2)] If there exists a positive integer $s$ such that $N|(p^s+1)$ and $\varphi(p^{-2s})\neq r^{2m}$, then $C$ is an $[2r^m, r^m]$ LCD code over $\mathbb{F}_q$.
      %In particular, if $\varphi(q^{-2})\neq 1$, then $C$ is a $[2r^m, r^m]$ LCD code over $\mathbb{F}_q$.
\end{enumerate}

\end{thm}
\begin{proof}
It follows from (\ref{eqn5}) that all eigenvalues of $PP^T$ are $0$ when $a=0$ and $\varphi(-1)(g(\varphi,\chi_a))^2$ when $a\in \mathbb{F}_{r^m}^*.$ According to Lemma \ref{lemm1}, we just have to prove that $\varphi(-1)(g(\varphi,\chi_a))^2\neq -1$ for any $a\in \mathbb{F}_{r^m}^*.$ Assume on the contrary that there exists $a\in \F_{r^m}^*$ such that $\varphi(-1)(g(\varphi,\chi_a))^2\\=-1$.

(1) If there exists a positive integer $s$ such that $N|(p^s-1)$, we get $\varphi^{p^s}=\varphi$. Then
\begin{eqnarray*}
% \nonumber to remove numbering (before each equation)
  && (\varphi(-1)(g(\varphi,\chi_a))^2)^{p^s}=(-1)^{p^s}=-1 \\
  &\Longrightarrow& (\varphi(-1))^{p^s}(g(\varphi,\chi_a))^{p^s})^2=-1 \\
   &\Longrightarrow& \varphi^{p^s}(-1)(g(\varphi^{p^s},\chi_{ap^s}))^2=-1 \\
   &\Longrightarrow& \varphi(-1)(g(\varphi,\chi_{ap^s}))^2=-1\\
   &\Longrightarrow& \varphi(-1)(\overline{\varphi}(p^s)g(\varphi,\chi_{a}))^2=-1\\
    &\Longrightarrow& \varphi(-1)(g(\varphi,\chi_{a}))^2=-(\overline{\varphi}(p^s))^{-2}=-\varphi(p^{2s}).
\end{eqnarray*}
Combined $\varphi(-1)(g(\varphi,\chi_a))^2=-1$ with $\varphi(-1)(g(\varphi,\chi_a))^2=-\varphi(p^{2s})$, we get $\varphi(p^{2s})=1$ which is a contradiction. Hence, $\varphi(-1)(g(\varphi,\chi_a))^2\neq -1$ for any $a\in \mathbb{F}_{r^m}^*.$

(2) If there exists a positive integer $s$ such that $N|(p^s+1)$, we get $\varphi^{p^s}=\varphi^{-1}$. Then
\begin{eqnarray*}
% \nonumber to remove numbering (before each equation)
  && (\varphi(-1)(g(\varphi,\chi_a))^2)^{p^s}=(-1)^{p^s}=-1\\
 % &\Longrightarrow& (\varphi(-1))^{p^s}(g(\varphi,\chi_a))^{p^s})^2=-1 \\
  &\Longrightarrow& \varphi^{p^s}(-1)(g(\varphi^{p^s},\chi_{ap^s}))^2=-1 \\
  &\Longrightarrow& \varphi^{-1}(-1)(g(\varphi^{-1},\chi_{ap^s}))^2=-1\\
  &\Longrightarrow& \varphi^{-1}(-1)(\overline{\varphi^{-1}}(p^s)\overline{\varphi^{-1}}(-1)g(\varphi^{-1},\chi_{-a}))^2=-1\\
  &\Longrightarrow& \varphi^{-1}(-1)(g(\varphi^{-1},\chi_{-a}))^2=-(\varphi^{-1}(p^{-s})\varphi^{-1}(-1))^{-2}=-\varphi(p^{-2s})\\
  &\Longrightarrow& \varphi^{-1}(-1)(\overline{g(\varphi,\chi_{a})})^2=-\varphi(p^{-2s}).
\end{eqnarray*}
Combined $\varphi(-1)(g(\varphi,\chi_a))^2=-1$ with $\varphi^{-1}(-1)(\overline{g(\varphi,\chi_{a})})^2=-\varphi(p^{-2s})$, we get
\begin{eqnarray*}
% \nonumber to remove numbering (before each equation)
  && \varphi(-1)(g(\varphi,\chi_a))^2\varphi^{-1}(-1)(\overline{g(\varphi,\chi_{a})})^2=\varphi(p^{-2s}) \\
  &\Longrightarrow& (g(\varphi,\chi_a)\overline{g(\varphi,\chi_{a})})^2=\varphi(p^{-2s})\\
   &\Longrightarrow& r^{2m}=\varphi(p^{-2s})~{\rm by~Lemma~\ref{lem2}},
\end{eqnarray*}
which is a contradiction. Therefore, $\varphi(-1)(g(\varphi,\chi_a))^2\neq -1$ for any $a\in \mathbb{F}_{r^m}^*.$

To sum up, $-1$ is not an eigenvalue of the matrix $PP^T$. By Lemma \ref{lemm1}, $C$ is an $[2r^m, r^m]$ LCD code over $\F_q$. This finishes the proof of the theorem.
\end{proof}
Two concrete examples with respect to Theorem \ref{thm22} are given as follows.
%In the following, we give an example to illustrate Theorem \ref{thm22}. In addition, the computations are carried out using Magma software \cite{B}.
\begin{ex}Let $r=7, m=1, N=3,p=2$ and $q=4$. Let $\mathbb{F}_4^*=\langle \beta\rangle$, where $\beta$ is a fixed primitive element of $\mathbb{F}_4^*.$ It is easy to check that $q,r,N$ satisfy the conditions in Theorem \ref{thm22}(1)(2). Then $C$ is a quaternary $[14,7,5]$ LCD code with generator matrix $[I_{7}, P]$, where
\begin{equation*}
{P}=\left(
  \begin{array}{ccccccccc}
    0 &  1 &\beta^2&  \beta&   \beta& \beta^2&   1 \\
   1 &  0&   1& \beta^2&   \beta &  \beta& \beta^2  \\
   \beta^2 &  1 &  0  & 1& \beta^2&   \beta &  \beta \\
   \beta& \beta^2&   1 &  0 &  1& \beta^2&   \beta\\
    \beta &  \beta& \beta^2 &  1 &  0 &  1& \beta^2\\
    \beta^2 &  \beta & \beta& \beta^2 &  1 &  0 &  1\\
     1& \beta^2 &  \beta &  \beta &\beta^2 &  1 &  0
  \end{array}
\right),
\end{equation*}which is almost optimal in the sense that the minimum distance of the optimal quaternary linear code with the length $14$ and the dimension $7$ is $6$ by the online Database \cite{G}. Moreover, the dual code of $C$ has parameters $[14,7,5]$, which is also almost optimal.
\end{ex}

\begin{ex}\label{ex2}Let $r=3, m=1, N=2$ and $q=p=5$. It is easy to check that $q,r,N$ satisfy the conditions in Theorem \ref{thm22}(2). Then $C$ is an $[6,3,3]$ LCD code over $\F_{5}$ with generator matrix $[I_{3}, P]$, where
\begin{equation*}
{P}=\left(
  \begin{array}{ccccccccc}
    0& 1& -1 \\
   -1& 0& 1  \\
   1& -1& 0 \\
  \end{array}
\right),
\end{equation*}which is almost optimal in the sense that the minimum distance of the optimal $5$-ary linear code with the length $6$ and the dimension $3$ is $4$ by the online Database \cite{G}. Moreover, the dual code of $C$ has parameters $[6,3,3]$, which is also almost optimal.
\end{ex}
In view of Theorem \ref{thm22}, since the sufficient condition is abstract for a linear code to be an LCD code, we present a concrete result as corollary in the following.
\begin{coro}Let $r$ be a prime number and $m$ be a positive integer. Assume that $N>1$ is a positive integer and $N|(r^m-1)$. Let $q$ be a power of prime $p$ and $(p,r)=1$. Assume that $N|(q-1)$.
%he $r^m\times r^m$ matrix $P=(p_{ij})\in M_{r^m}(\mathbb{F}_q)$ is defined as (\ref{mat}).
Let $C:=C_{(\varphi,0)}$ be the linear code over $\mathbb{F}_q$ with generator matrix $[I_{r^m}, P]$. Then we have the following.
\begin{enumerate}
  \item [(1)] If there exists a positive integer $s$ such that $N|(p^s-1)$ and $p^{\frac{2s(r^m-1)}{N}}\not\equiv 1~({\rm mod~}r)$, then $C$ is an $[2r^m, r^m]$ LCD code over $\mathbb{F}_q$.
      %In particular, if $\varphi(q^{2})\neq 1$, then $C$ is a $[2r^m, r^m]$ LCD code over $\mathbb{F}_q$.
  \item [(2)] If there exists a positive integer $s$ such that $N|(p^s+1)$ and $r^{2mN}\not\equiv 1~({\rm mod~}p)$, then $C$ is an $[2r^m, r^m]$ LCD code over $\mathbb{F}_q$.
      %In particular, if $\varphi(q^{-2})\neq 1$, then $C$ is a $[2r^m, r^m]$ LCD code over $\mathbb{F}_q$.
\end{enumerate}
\begin{proof}Compared with the conditions of Theorem \ref{thm22}, we just have to prove that (1) if $p^{\frac{2s(r^m-1)}{N}}\not\equiv 1~({\rm mod~}r)$, then
$\varphi(p^{2s})\neq 1$ and (2) if $r^{2mN}\not\equiv 1~({\rm mod~}p)$, then $\varphi(p^{-2s})\neq r^{2m}$, respectively.

(1) Assume on the contrary that $\varphi(p^{2s})=1$, then $p^{2s}\in \ker(\varphi)$. Hence, $p^{2s}\in \langle \alpha^N\rangle$. Since ${\rm ord}(\alpha^N)=\frac{r^m-1}{N}$, we have $p^{2s\cdot\frac{r^m-1}{N}}\equiv 1~({\rm mod}~r)$, it is a contradiction.

(2) Assume on the contrary that $\varphi(p^{-2s})=r^{2m}$, then $(\varphi(p^{-2s}))^N=r^{2mN}$. Hence, $r^{2mN}\equiv 1~({\rm mod}~p)$, it is a contradiction. This completes the proof.
\end{proof}
\end{coro}
\hspace*{-0.5cm}\emph{\textbf{Construction B.} Define $\rho(0)=v$. We then obtain a $r^m\times r^m$ matrix $P=(p_{ij})$ by $$p_{ij}=\rho(x_j-x_i),$$ which is defined as (\ref{mat}).
%It follows from (\ref{den4}) that all eigenvalues of $PP^T$ are given by
%\begin{equation}\label{eqn6}
%\lambda_a=\begin{cases}
%\emph{ }v^2,  ~~~~~~~~~~~~~~~~~~~~~~~~~~~~~~~~~~~~~~~~~~~~~~~~~~~~~~~~~~~~~~~~~{\rm if}~a=0;\\
 %  \emph{ }v^2+(1+\varphi(-1)){\overline{\varphi}(a)}g(\varphi,\chi_1)v+\varphi(-1)({\overline{\varphi}(a)}g(\varphi,\chi_1))^2, ~{\rm if}~a\in \mathbb{F}_{r^m}^*. \\
%\end{cases}
%\end{equation}}
For any $a\in \F_{r^m}$, we define $f_a: \F_q\longrightarrow \overline{\F}_{q}$ by
\begin{equation}\label{eqn7}
f_a(x)=\begin{cases}
\emph{ }x^2,  ~~~~~~~~~~~~~~~~~~~~~~~~~~~~~~~~~~~~~~~~~~~~~~~~~~~~~~~~~~~~~~~~{\rm if}~a=0;\\
   \emph{ }x^2+(1+\varphi(-1)){\overline{\varphi}(a)}g(\varphi,\chi_1)x+\varphi(-1)({\overline{\varphi}(a)}g(\varphi,\chi_1))^2, {\rm if}~a\in \mathbb{F}_{r^m}^*. \\
\end{cases}
\end{equation}
It follows from (\ref{den4}) that all eigenvalues of $PP^T$ are given by $$\lambda_a:=f_a(v)$$ for all $a\in \F_{r^m}$.}

In order to construct LCD codes over $\F_q$, we hope that there exists $v\in \mathbb{F}_q$ satisfying $f_a(v)\neq-1$ for any $a\in \F_{r^m}$. Hence, we present a lemma as follows.
\begin{lem}\label{lemm2} Let the symbols be the same as above. If $q> 2(N+1)$, then there exists $v\in \mathbb{F}_q$ satisfying $\lambda_a:=f_a(v)\neq-1$ for any $a\in \F_{r^m}$.
\end{lem}
\begin{proof}
Since the order of $\varphi$ is $N$, the set $\{f_a(x): a\in \F_{r^m}\}$ has at most $N+1$ distinct polynomials. For any $a\in \F_{r^m},$ $f_a(x)=-1$ has at most two solutions in $\F_q$.
Theorefore, all these equations in $\{f_a(x)=-1: a\in \F_{r^m}\}$ have at most $2(N+1)$ solutions over $\F_q$.
Since $q> 2(N+1)$, there exists an element $v\in \F_q$ such that $v$ is not a solution of any equation $f_a(x)=-1$, i.e., there exists $v\in \F_q$ satisfying ${\lambda}_a=f_a(v)\neq -1$ for any $a\in \F_{r^m}$.
\end{proof}
Based on the discussion above, we can easily get the following theorem.
\begin{thm}\label{thm212}
Let $r$ be a prime number and $m$ be a positive integer. Assume that $N>1$ is a positive integer and $N|(r^m-1)$. Let $q$ be a power of prime $p$ and $(p,r)=1$. Assume that $N|(q-1)$.
%he $r^m\times r^m$ matrix $P=(p_{ij})\in M_{r^m}(\mathbb{F}_q)$ is defined as (\ref{mat}).
Let $C:=C_{(\varphi,v)}$ be the linear code over $\mathbb{F}_q$ with generator matrix $[I_{r^m}, P]$. If $q> 2(N+1)$, then {there exists $v\in \F_q$ such that} $C$ is an $[2r^m, r^m]$ LCD code over $\mathbb{F}_q$.
\end{thm}
\begin{proof} By Lemmas \ref{lemm1} and \ref{lemm2}, we can easily obtain the desired results. So we omit
the detail here.
\end{proof}
Next, we present an example to explain the result of Theorem \ref{thm212}.
\begin{ex}\label{exa3}Let $r=2, m=2, N=3,p=5$ and $q=25$. Let $\mathbb{F}_{25}^*=\langle \beta\rangle$, where $\beta$ is a fixed primitive element of $\mathbb{F}_{25}^*.$ Taking $v=\beta^2$. It is easy to check that $q,r,N$ satisfy the conditions in Theorem \ref{thm212}. Then $C$ is an $[8,4,4]$ LCD code over $\F_{25}$ with generator matrix $[I_{4}, P]$, where
\begin{equation*}
{P}=\left(
  \begin{array}{ccccccccc}
     \beta^2& \beta^{16}  &\beta^8  &1 \\
   \beta^{16}  &       \beta^2   &      1  &\beta^{8}  \\
   \beta^{8}   &   1       &  \beta^2 & \beta^{16}\\
     1   &\beta^{8} & \beta^{16}  &       \beta^2\\
  \end{array}
\right),
\end{equation*}which is an almost MDS code.
 %in the sense that the minimum distance of the optimal linear code with length $8$ and dimension $4$ is $5$ according to the definition of MDS codes.
Moreover, the dual code of $C$ has parameters $[8,4,4]$, which is also an almost MDS code.
\end{ex}

\section{The constructions of linear codes with one-dimensional hull}
%Before , we first give the sufficient conditions for linear codes with one-dimensional hull via their generator matrices in the standard form as follows.

In this section, we present the constructions of linear codes with one-dimensional hull by using the two homomorphisms (\ref{eq1}) and (\ref{eq2}). In order to construct linear codes with one-dimensional hull over $\F_q$, we hope that there exists $v\in \mathbb{F}_q$ satisfying $\lambda_0=v^2=-1$ and $\lambda_a\neq -1$ for any $a\in \F_{r^m}$ by Lemma \ref{lem4}. Let $q$ be a power of a prime $p$. In what follows, we shall consider the construction dividing into two cases $p=2$ and $p\geq 3$.
\subsection{The case $p=2$}
%\hspace*{-0.5cm}\emph{\textbf{Case 1: $p=2$.}}
Define $\rho(0)=v=1$. Then $v^2=1=-1$. We then obtain a $r^m\times r^m$ matrix $P=(p_{ij})$ by $$p_{ij}=\rho(x_j-x_i),$$ which is defined as (\ref{mat}). It follows from (\ref{den4}) that all eigenvalues of $PP^T$ are given by
\begin{equation}\label{eqn55}
\lambda_a=\begin{cases}
\emph{ }-1,  ~~~~~~~~~~~~~~~~~~~~~{\rm if}~a=0;\\
   \emph{ }-1+(g(\varphi,\chi_a))^2, ~~~{\rm if}~a\in \mathbb{F}_{r^m}^*. \\
\end{cases}
\end{equation}
\hspace*{0.5cm}Therefore, we present the following theorem.
\begin{thm}\label{thm11} Let $r$ be an odd prime number and $m$ be a positive integer. Assume that $N>1$ is a positive integer and $N|(r^m-1)$. Let $q$ be a power of $p=2$ and $N|(q-1)$.
%The $r^m\times r^m$ matrix $P=(p_{ij})\in M_{r^m}(\mathbb{F}_q)$ is defined as (\ref{mat}).
Let $C:=C_{(\varphi,1)}$ be the linear code over $\mathbb{F}_q$ with generator matrix $[I_{r^m}, P]$. Then $C$ is a $[2r^m, r^m]$ linear code over $\mathbb{F}_q$ with one-dimensional hull.
\end{thm}
\begin{proof} It follows from (\ref{eqn55}) that all eigenvalues of $PP^T$ are $-1$ when $a=0$ and $-1+(g(\varphi,\chi_a))^2$ when $a\in \mathbb{F}_{r^m}^*.$ By using Lemma \ref{lem4}, we just have to prove that $-1+(g(\varphi,\chi_a))^2\neq -1$ for any $a\in \mathbb{F}_{r^m}^*.$ Note that the result $g(\varphi,\chi_a)\overline{g(\varphi,\chi_a)}=r^m$ for any $a\in \mathbb{F}_{r^m}^*$ from Lemma \ref{lem2}. Then $g(\varphi,\chi_a)\neq 0$ and $\overline{g(\varphi,\chi_a)}\neq 0$ for any $a\in \mathbb{F}_{r^m}^*$. Hence, $(g(\varphi,\chi_a))^2\neq 0$ and $-1+(g(\varphi,\chi_a))^2\neq -1$ for any $a\in \mathbb{F}_{r^m}^*$. The desired conclusion then follows.
\end{proof}
Here, we give a concrete example as follows.
\begin{ex} Let $r=13, m=1, N=3, p=2$ and $q=4$. Let $\mathbb{F}_4^*=\langle \beta\rangle$, where $\beta$ is a fixed primitive element of $\mathbb{F}_4^*.$ It is easy to check that $q,r,N$ satisfy the conditions in Theorem \ref{thm11}. Then $C$ is a $[26,13,8]$ linear code over $\F_{4}$ with one-dimensional hull and its generator matrix $[I_{13}, P]$, where
\begin{equation*}
{P}=\left(
  \begin{array}{cccccccccccccc}
    1& 1 &\beta &\beta &\beta^2 &1 &\beta^2 &\beta^2 &1 &\beta^2 &\beta& \beta& 1\\
    1& 1 &1 &\beta& \beta& \beta^2 &1& \beta^2& \beta^2 &1 &\beta^2& \beta& \beta \\
     \beta& 1& 1& 1& \beta &\beta& \beta^2& 1 &\beta^2& \beta^2& 1& \beta^2& \beta \\
    \beta &\beta &1 &1 &1& \beta &\beta& \beta^2 &1& \beta^2 &\beta^2 &1 &\beta^2 \\
    \beta^2& \beta& \beta& 1 &1 &1& \beta &\beta &\beta^2& 1 &\beta^2& \beta^2& 1\\
    1 &\beta^2& \beta& \beta &1& 1& 1 &\beta &\beta& \beta^2& 1 &\beta^2& \beta^2\\
    \beta^2 &1 &\beta^2& \beta& \beta& 1& 1& 1& \beta& \beta &\beta^2 &1 &\beta^2\\
    \beta^2& \beta^2& 1& \beta^2& \beta& \beta& 1& 1 &1& \beta& \beta &\beta^2& 1\\
    1& \beta^2& \beta^2& 1& \beta^2& \beta& \beta &1 &1& 1& \beta &\beta& \beta^2\\
     \beta^2& 1 &\beta^2& \beta^2& 1 &\beta^2& \beta &\beta &1 &1& 1& \beta& \beta\\
     \beta &\beta^2 &1& \beta^2& \beta^2& 1& \beta^2 &\beta& \beta &1 &1& 1& \beta\\
      \beta& \beta& \beta^2 &1& \beta^2& \beta^2& 1& \beta^2& \beta &\beta &1 &1 &1\\
       1 &\beta& \beta& \beta^2& 1& \beta^2& \beta^2 &1 &\beta^2& \beta& \beta& 1 &1\\
  \end{array}
\right).
\end{equation*}
Moreover, the hull of $C$ is a $[26,1,26]$ cyclic code over $\F_{4}$ with generator
matrix
\begin{equation*}
\left(
  \begin{array}{cccccccccccccccccccccccccccccccc}
     1& 1& 1& 1 &1 &1& 1 &1& 1& 1&1& 1& 1&1& 1&1&1&1&1&1&1&1&1&1&1&1\\
  \end{array}
\right).
\end{equation*}
\end{ex}
\subsection{The case $p\geq 3$}
%\hspace*{-0.5cm}\emph{\textbf{Case 2: $p\geq 3$.}}
In this subsection, we let $\mathbb{F}_q^*=\langle \beta\rangle$, where $\beta$ is a fixed primitive element of $\mathbb{F}_q^*.$ Assume that $4|(q-1)$.

Define $\rho(0)=v=\beta^{\frac{q-1}{4}}$. Then $v^2=(\beta^\frac{q-1}{4})^2=\beta^\frac{q-1}{2}=-1$. We then obtain a $r^m\times r^m$ matrix $P=(p_{ij})$ by $$p_{ij}=\rho(x_j-x_i),$$ which is defined as (\ref{mat}). In addition, $\varphi(-1)=\varphi(\alpha^\frac{r^m-1}{2})=u^\frac{r^m-1}{2}=(\beta^\frac{q-1}{N})^{\frac{r^m-1}{2}}=(\beta^\frac{q-1}{2})^\frac{r^m-1}{N}=(-1)^\frac{r^m-1}{N}$. When $\frac{r^m-1}{N}$ is odd, $\varphi(-1)=-1$; when $\frac{r^m-1}{N}$ is even, $\varphi(-1)=1$.

Combined with (\ref{den4}), when $\frac{r^m-1}{N}$ is odd, we have
\begin{equation}\label{eq4}
\lambda_a=\begin{cases}
\emph{ }-1,  ~~~~~~~~~~~~~~~~~~~{\rm if}~a=0;\\
   \emph{ }-1-(g(\varphi,\chi_a))^2, ~{\rm if}~a\in \mathbb{F}_{r^m}^*; \\
\end{cases}
\end{equation}
when $\frac{r^m-1}{N}$ is even, we get
\begin{equation}\label{eq5}
\lambda_a=\begin{cases}
\emph{ }-1,  ~~~~~~~~~~~~~~~~~~~~~~~~~~~~~~~~~~{\rm if}~a=0;\\
   \emph{ }-1+(2v+g(\varphi,\chi_a))g(\varphi,\chi_a), ~{\rm if}~a\in \mathbb{F}_{r^m}^*. \\
\end{cases}
\end{equation}

Collecting all discussions above, we first present the sufficient conditions for constructing linear codes with one-dimensional hull when $\frac{r^m-1}{N}$ is odd.
%obtain the following result.
\begin{thm}\label{thm1} Let $r$ be a prime number and $m$ be a positive integer. Assume that $N>1$ is a positive integer and $N|(r^m-1)$. Let $q$ be a power of prime $p$ and $(p,r)=1$. Assume that $N|(q-1)$ and $4|(q-1)$.
%The $r^m\times r^m$ matrix $P=(p_{ij})\in M_{r^m}(\mathbb{F}_q)$ is defined as (\ref{mat}).
Let $C:=C_{(\varphi, \beta^{\frac{q-1}{4}})}$ be the linear code over $\mathbb{F}_q$ with generator matrix $[I_{r^m}, P]$. When $\frac{r^m-1}{N}$ is odd, $C$ is a $[2r^m, r^m]$ linear code over $\mathbb{F}_q$ with one-dimensional hull.
\end{thm}
\begin{proof} The proof is similar to that of Theorem \ref{thm11} and omitted here.
\end{proof}
%\begin{proof} It follows from (\ref{eq4}) that all eigenvalues of $PP^T$ are $-1$ when $a=0$ and $-1-(g(\varphi,\chi_a))^2$ when $a\in \mathbb{F}_{r^m}^*.$ According to Lemma \ref{lem4}, we just have to prove that $-1-(g(\varphi,\chi_a))^2\neq -1$ for any $a\in \mathbb{F}_{r^m}^*.$ Note that the result $g(\varphi,\chi_a)\overline{g(\varphi,\chi_a)}=r^m$ for $a\in \mathbb{F}_{r^m}^*$ from Lemma \ref{lem2}. Then $g(\varphi,\chi_a)\neq 0$ and $\overline{g(\varphi,\chi_a)}\neq 0$ for $a\in \mathbb{F}_{r^m}^*$. Hence, $(g(\varphi,\chi_a))^2\neq 0$ and $-1-(g(\varphi,\chi_a))^2\neq -1$ for $a\in \mathbb{F}_{r^m}^*$. The desired conclusion then follows.
%\end{proof}
\begin{ex}\label{exa} Let $r=3, m=2, N=8, p=7$ and $q=49$. Let $\mathbb{F}_{49}^*=\langle \beta\rangle$, where $\beta$ is a fixed primitive element of $\mathbb{F}_{49}^*.$ It is easy to check that $q,r,N$ satisfy the conditions in Theorem \ref{thm1}. Then $C$ is a $[18,9,8]$ linear code over $\F_{49}$ with one-dimensional hull and its generator matrix $[I_{9}, P]$, where
\begin{equation*}
{P}=\left(
  \begin{array}{cccccccccc}
    \beta^{12}& \beta^{42}& \beta^6& \beta^{30}& 1& \beta^{36}& \beta^{18}& \beta^{12} &6\\
    \beta^{18}& \beta^{12}& 1& \beta^{12} &\beta^{36}& \beta^6& \beta^{42}& 6 &\beta^{30} \\
    \beta^{30} &6& \beta^{12} &\beta^6& \beta^{18}& \beta^{42}& \beta^{12}& 1 &\beta^{36} \\
    \beta^6& \beta^{36}& \beta^{30}& \beta^{12}& \beta^{12}& 6& 1& \beta^{18} &\beta^{42} \\
    6& \beta^{12}& \beta^{42}& \beta^{36}& \beta^{12}& \beta^{18}& \beta^{30}& \beta^6& 1\\
    \beta^{12}& \beta^{30}& \beta^{18}& 1& \beta^{42}& \beta^{12}& 6& \beta^{36} &\beta^6\\
    \beta^{42}& \beta^{18}& \beta^{36}& 6 &\beta^6& 1& \beta^{12}& \beta^{30}& \beta^{12}\\
    \beta^{36}& 1 &6& \beta^{42}& \beta^{30}& \beta^{12}& \beta^6 &\beta^{12} &\beta^{18}\\
    1& \beta^6 &\beta^{12}& \beta^{18}& 6& \beta^{30} &\beta^{36}& \beta^{42}& \beta^{12}
  \end{array}
\right).
\end{equation*}
Moreover, the hull of $C$ is a $[18,1,18]$ quasi-cyclic code of index $2$ over $\F_{49}$ with generator
matrix
\begin{equation*}
\left(
  \begin{array}{cccccccccccccccccc}
     1& 1& 1& 1 &1 &1& 1 &1& 1& \beta^{12}&\beta^{12}& \beta^{12}& \beta^{12}& \beta^{12}& \beta^{12}& \beta^{12}&
    \beta^{12}& \beta^{12}\\
  \end{array}
\right).
\end{equation*}
Compared with Example 2(2) in \cite{LZ}, the linear code $C$ over $\F_{49}$ with one-dimensional hull we obtained has better parameters than its parameters. In other words, the linear code $C$ of the length $18$ with the dimension $9$ has the minimal distance $8$, while the linear code $C$ of the length $18$ with the dimension $9$ in \cite[ Example 2(2)]{LZ} has the minimal distance $7$. That is to say, the linear code $C$ over $\F_{49}$ with one-dimensional hull we obtained is also considered new.
\end{ex}

\begin{ex}\label{exa2} Let $r=7, m=1, N=6, p=5$ and $q=25$. Let $\mathbb{F}_{25}^*=\langle \beta\rangle$, where $\beta$ is a fixed primitive element of $\mathbb{F}_{25}^*.$ It is easy to check that $q,r,N$ satisfy the conditions in Theorem \ref{thm1}. Then $C$ is a $[14,7,7]$ linear code over $\F_{25}$ with one-dimensional hull and its generator matrix $[I_{7}, P]$, where
\begin{equation*}
{P}=\left(
  \begin{array}{cccccccccc}
     2&       1&  \beta^8 & \beta^4& \beta^{16}& \beta^{20} &      4\\
      4 &      2  &     1 & \beta^8&  \beta^4& \beta^{16}& \beta^{20}&\\
     \beta^{20} &      4&       2 &      1 & \beta^8&  \beta^4 &\beta^{16}& \\
   \beta^{16} &\beta^{20} &      4  &     2&       1 & \beta^8&  \beta^4 \\
    \beta^4& \beta^{16}& \beta^{20} &      4  &     2  &     1 & \beta^8\\
    \beta^8 & \beta^4& \beta^{16}& \beta^{20} &      4  &     2 &      1\\
        1 & \beta^8 & \beta^4 &\beta^{16} &\beta^{20} &      4 &      2
  \end{array}
\right),
\end{equation*}which is an almost MDS code.
% In addition, the dual code of $C$ has parameters $[14,7,7]$, which is also an almost MDS code.
 %in the sense that the minimum distance of the optimal linear code with length $8$ and dimension $4$ is $5$ according to the definition of MDS codes.
Moreover, the hull of $C$ is a $[14,1,14]$ quasi-cyclic code of index $2$ over $\F_{25}$ with generator
matrix
\begin{equation*}
\left(
  \begin{array}{cccccccccccccccccc}
     1& 1& 1& 1 &1 &1& 1 &2& 2&2&2& 2& 2& 2\\
  \end{array}
\right).
\end{equation*}

\end{ex}

Next, we turn to the sufficient conditions for constructing linear codes with one-dimensional hull when $\frac{r^m-1}{N}$ is even.
\begin{thm}\label{thm2} Let $r$ be a prime number and $m$ be a positive integer. Assume that $N>1$ is a positive integer and $N|(r^m-1)$. Let $q$ be a power of prime $p$ and $(p,r)=1$. Assume that $N|(q-1)$ and $4|(q-1)$.
%Let $\mathbb{F}_q^*=\langle \beta\rangle$ and $v=\beta^\frac{q-1}{4}$. The $r^m\times r^m$ matrix $P=(p_{ij})\in M_{r^m}(\mathbb{F}_q)$ is defined as (\ref{mat}).
Let $C:=C_{(\varphi, \beta^{\frac{q-1}{4}})}$ be the linear code over $\mathbb{F}_q$ with generator matrix $[I_{r^m}, P]$. When $\frac{r^m-1}{N}$ is even and $2v+g(\varphi, \chi_a)\neq 0$ for all $a\in \mathbb{F}_{r^m}^*$, $C$ is a $[2r^m, r^m]$ linear code over $\mathbb{F}_q$ with one-dimensional hull.
%Then we have the following.
%\begin{enumerate}
 % \item [(1)] When $p=2$, $C$ is a $[2r^m, r^m]$ linear code over $\mathbb{F}_q$ with one-dimensional hull.
%  \item [(2)]
%\end{enumerate}
\end{thm}
\begin{proof} It follows from (\ref{eq5}) that all eigenvalues of $PP^T$ are $-1$ when $a=0$ and $-1+(2v+g(\varphi,\chi_a))g(\varphi,\chi_a)$ when $a\in \mathbb{F}_{r^m}^*.$ According to Lemma \ref{lem4}, we just have to prove that $-1+(2v+g(\varphi,\chi_a))g(\varphi,\chi_a)\neq -1$ for all $a\in \mathbb{F}_{r^m}^*.$

%When $p=2$, $-1+(2v+g(\varphi,\chi_a))g(\varphi,\chi_a)=-1+(g(\varphi,\chi_a))^2\neq -1$ for $a\in \mathbb{F}_{r^m}^*$ by the similar proof of Theorem \ref{thm1}.
By utilizing Lemma \ref{lem2} and the proof of Theorem \ref{thm11}, we obtain that $g(\varphi, \chi_a)\neq 0$ for any $a\in \mathbb{F}_{r^m}^*$. When $2v+g(\varphi, \chi_a)\neq 0$ for all $a\in \mathbb{F}_{r^m}^*$, we have $-1+(2v+g(\varphi,\chi_a))g(\varphi,\chi_a)\neq -1$ for all $a\in \mathbb{F}_{r^m}^*.$

Therefore, the matrix ${P}{P}^T$ has an eigenvalue $-1$ with multiplicity $1$. It then follows from Lemma \ref{lem4} that $C$ is a $[2r^m, r^m]$ linear code over $\mathbb{F}_q$ with one-dimensional hull.
\end{proof}
In Theorem \ref{thm2}, the condition ``$2v+g(\varphi, \chi_a)\neq 0$ for all $a\in \mathbb{F}_{r^m}^*$" is not very straightforward. Hence, we will present the following corollary as a concrete result.
\begin{coro}\label{coro}
%Let the symbols be the same as those in Theorem \ref{thm2}.
Let $r$ be a prime number and $m$ be a positive integer. Assume that $N>1$ is a positive integer and $N|(r^m-1)$. Let $q$ be a power of odd prime $p$ and $(p,r)=1$. Assume that $N|(q-1)$ and $4|(q-1)$.
%The $r^m\times r^m$ matrix $P=(p_{ij})\in M_{r^m}(\mathbb{F}_q)$ is defined as (\ref{mat}).
Let $C:=C_{(\varphi, \beta^{\frac{q-1}{4}})}$ be the linear code over $\mathbb{F}_q$ with generator matrix $[I_{r^m}, P]$. Let $\frac{r^m-1}{N}$ be even. If $\varphi(q)\neq 1$, then $C$ is a $[2r^m, r^m]$ linear code over $\mathbb{F}_q$ with one-dimensional hull.
\end{coro}
\begin{proof} Since $\frac{r^m-1}{N}$ is even and it follows from (\ref{eq5}) that all eigenvalues of $PP^T$ are $-1$ when $a=0$ and $-1+(2v+g(\varphi,\chi_a))g(\varphi,\chi_a)$ when $a\in \mathbb{F}_{r^m}^*.$ Suppose that $2v+g(\varphi,\chi_a)=0$ for some $a\in \mathbb{F}_{r^m}^*$. Then $g(\varphi,\chi_a)=-2v\in \mathbb{F}_{p}$ when $p\equiv 1~({\rm mod}~4)$ and $g(\varphi,\chi_a)=-2v\in \mathbb{F}_{p^2}$ when $p\equiv 3~({\rm mod}~4)$.
 %by $v=\beta^\frac{q-1}{4}\in \mathbb{F}_{p^2}$.
In addition,
\begin{eqnarray*}
% \nonumber to remove numbering (before each equation)
  (g(\varphi,\chi_a))^q &=& \left(\sum\limits_{x\in \mathbb{F}_{r^m}^*}\varphi(x)\chi_a(x)\right)^q \\
  &=& g(\varphi^q,\chi_{aq}) \\
  &=& g(\varphi,\chi_{aq})\\
  &=& \varphi(q^{-1})g(\varphi,\chi_{a})\\
  &=& \varphi(q)^{-1}g(\varphi,\chi_{a})
\end{eqnarray*}by $N|(q-1)$ and Section 3(\ref{eqn8}). If $\varphi(q)\neq 1,$ then $(g(\varphi,\chi_{a}))^q\neq g(\varphi,\chi_{a})$, i.e., $g(\varphi,\chi_{a})\notin \mathbb{F}_{q}$.

When $p\equiv 1~({\rm mod}~4)$, $\mathbb{F}_{p}\subseteq \mathbb{F}_q$, which implies that $g(\varphi,\chi_{a})\notin \mathbb{F}_{p}.$ It is a contradiction.

When $p\equiv 3~({\rm mod}~4)$, $\mathbb{F}_{p^2}\subseteq \mathbb{F}_q$ by $4|(q-1)$, which implies that $g(\varphi,\chi_{a})\notin \mathbb{F}_{p^2}.$
%Since $\mathbb{F}_{p^2}\subseteq \mathbb{F}_q$, we get $g(\varphi,\chi_{a})\notin \mathbb{F}_{p^2}.$
It is a contradiction.

Hence, $2v+g(\varphi,\chi_a)\neq0$. By using Lemma \ref{lem2}, we obtain that $g(\varphi,\chi_a)\neq 0.$ Then $-1+(2v+g(\varphi,\chi_a))g(\varphi,\chi_a)\neq -1$ for all $a\in \mathbb{F}_{r^m}^*$. Thus the matrix ${P}{P}^T$ has an eigenvalue $-1$ with multiplicity $1$. It then follows from Lemma \ref{lem4} that  the desired result then follows.
\end{proof}
We now employ Corollary \ref{coro} to present a example as follows.
\begin{ex} Let $r=7, m=1, N=3, p=5$ and $q=25$. Let $\mathbb{F}_{25}^*=\langle \beta\rangle$, where $\beta$ is a fixed primitive element of $\mathbb{F}_{25}^*.$ It is easy to check that $q,r,N$ satisfy the conditions in Corollary \ref{coro}. Then $C$ is a $[14,7,6]$ linear code over $\F_{25}$ with one-dimensional hull and its generator matrix $[I_{7}, P]$, where
\begin{equation*}
{P}=\left(
  \begin{array}{ccccccccc}
    2& 1& \beta^{16}& \beta^8& \beta^8& \beta^{16}& 1\\
   1& 2& 1& \beta^{16}& \beta^8& \beta^8& \beta^{16} \\
   \beta^{16}& 1& 2& 1& \beta^{16}& \beta^8& \beta^8\\
   \beta^8& \beta^{16}& 1& 2& 1& \beta^{16}& \beta^8\\
   \beta^8& \beta^8& \beta^{16}& 1& 2& 1& \beta^{16} \\
   \beta^{16}& \beta^8& \beta^8& \beta^{16}& 1& 2& 1\\
   1& \beta^{16}& \beta^8& \beta^8& \beta^{16}& 1& 2
  \end{array}
\right).
\end{equation*}
Moreover, the hull of $C$ is a $[14,1,14]$ quasi-cyclic code of index $2$ over $\F_{25}$ with generator
matrix
\begin{equation*}
\left(
  \begin{array}{cccccccccccccccccc}
     1&1  &     1    &   1    &   1    &   1   &    1     &  2    &   2    &   2  &     2  &     2    &   2   &    2\\
  \end{array}
\right).
\end{equation*}
\end{ex}
Furthermore, some optimal or almost optimal LCD codes (resp. linear codes with one-dimensional hull) derived from Theorems \ref{thm22} and \ref{thm212} (resp. Theorems \ref{thm11}, \ref{thm1} and Corollary \ref{coro}) are listed in Table \ref{table} by Magma \cite{B}.
\begin{table}[h]
\caption{The list of \textbf{optimal or almost optimal} linear codes over small fields}
\label{table}
\setlength{\tabcolsep}{3pt}
\begin{tabular}{|p{110pt}|p{120pt}|p{30pt}|p{60pt}|p{80pt}|}
\hline & $r, m,N$ &$\mathbb{F}_q$& $[n,k,d]$& Theorems
\\\hline
&$r=13,m=1,N=3$&$\mathbb{F}_7$&$[26,13,9]^\star$&Theorem~\ref{thm22}(1)
\\&$r=13,m=1,N=4$&$\mathbb{F}_5$&$[26,13,9]^\star$&Theorem~\ref{thm22}(1)
\\&$r=17,m=1,N=8$&$\mathbb{F}_9$&$[34,17,12]^*$&Theorem~\ref{thm22}(1)
\\&$r=17,m=1,N=4$&$\mathbb{F}_5$&$[34,17,11]^*$&Theorem~\ref{thm22}(1)
\\&$r=5,m=1,N=2$&$\mathbb{F}_7$&$[10,5,5]^\star$&Theorem~\ref{thm22}(2)
\\& $r=7,m=1,N=2$&$\mathbb{F}_5$&$[14,7,6]^*$&Theorem~\ref{thm22}(2)
\\ {LCD codes}& $r=11,m=1,N=2$&$\mathbb{F}_7$&$[22,11,8]^\star$&Theorem~\ref{thm22}(2)
\\& $r=13,m=1,N=2$&$\mathbb{F}_5$&$[26,13,9]^\star$&Theorem~\ref{thm22}(2)
\\& $r=17,m=1,N=2$&$\mathbb{F}_7$&$[34,17,11]^\star$&Theorem~\ref{thm22}(2)
\\& $r=17,m=1,N=4$&$\mathbb{F}_9$&$[34,17,11]^\star$&Theorem~\ref{thm22}(2)
\\& $r=17,m=1,N=2$&$\mathbb{F}_5$&$[34,17,11]^*$&Theorem~\ref{thm22}(2)
\\&$r=3,m=1,N=2$&$\mathbb{F}_7$&$[6,3,3]^\star$&Theorem~\ref{thm212}
\\&$r=3,m=2,N=2$&$\mathbb{F}_7$&$[18,9,7]^\star$&Theorem~\ref{thm212}
\\&$r=5,m=1,N=2$&$\mathbb{F}_7$&$[10,5,5]^*$&Theorem~\ref{thm212}
\\&$r=11,m=1,N=2$&$\mathbb{F}_7$&$[22,11,8]^\star$&Theorem~\ref{thm212}
\\\hline
&$r=7,m=1,N=3$&$\mathbb{F}_4$&$[14,7,6]^*$&Theorem~\ref{thm11}
\\{Linear codes with}& $r=3,m=1,N=2$&$\mathbb{F}_5$&$[6,3,3]^\star$&Theorem~\ref{thm1}
\\{one-dimensional hull}&$r=7,m=1,N=2$&$\mathbb{F}_5$&$[14,7,6]^*$&Theorem~\ref{thm1}
\\&$r=11,m=1,N=2$&$\mathbb{F}_9$&$[22,11,8]^\star$&Theorem~\ref{thm1}
\\&$r=17,m=1,N=4$&$\mathbb{F}_9$&$[34,17,11]^\star$&Corollary~\ref{coro}
\\&$r=17,m=1,N=8$&$\mathbb{F}_9$&$[34,17,11]^\star$&Corollary~\ref{coro}
\\&$r=17,m=1,N=2$&$\mathbb{F}_5$&$[34,17,11]^*$&Corollary~\ref{coro}
\\\hline
\end{tabular}
\label{tab1}The codes with asterisk $(^*)$ have the property that linear codes are best known $q$-ary linear codes in \cite{B}, which is optimal. The codes with asterisk $(^\star)$ have the property that linear codes have better parameters according to the Database \cite{B}, which is almost optimal in the sense. For example, the linear code over $\F_9$ of the length $n=10$ with the dimension $k=5$ has the minimum distance $5$, while the code in the Database \cite{B} has the minimum distance $6$.
\end{table}
\begin{rem} In Table \ref{table}, optimal linear codes with one-dimensional hull in \cite[Section A]{LZ} can also be obtained by our construction methods when $N=2$ (see the first row, fourth row and fifth row of Table \ref{table}), which implies that our results contain partial results in \cite[Section A]{LZ}. When $N>2$, the linear codes are different from those in \cite[Section A]{LZ}. In addition, although the second and third low parameters of Table \ref{table} are the same, we verified by Magma that the two codes are not equivalent.
\end{rem}
\section{The minimum distance of the linear code $C_{(\varphi,v)}$}
In this section, we discuss the lower bound on the minimum distance of linear code $C:=C_{(\varphi,v)}$ with generator matrix $G=[I_{r^m}, P]$ defined in Section 3.

Assume that $q$ is a power of odd prime $p$ and $N=2$. Let $\mathbb{F}_{r^m}=\{x_i: 1\leq i\leq r^m\}$, where $x_1, \cdots, x_{\frac{r^m-1}{2}}$ are non-zero squares in $\mathbb{F}_{r^m}$, $x_{\frac{r^m+1}{2}}, \cdots, x_{r^m-1}$ are non-squares in $\mathbb{F}_{r^m}$ and $x_{r^m}=0$. From Section 3, we have $P\eta_a=\theta_a\eta_a$ for any $a\in \mathbb{F}_{r^m}$, where $\theta_a:=\sum\limits_{y\in  \mathbb{F}_{r^m}}\rho(y)\chi_a(y)$ and $\eta_a:=(\chi_a(x_1), \chi_a(x_2), \cdots, \chi_a(x_{r^m}))^T$.
Let $Q:=(\eta_{x_1}, \eta_{x_2}, \cdots, \eta_{x_{r^m}})$. Then
 \begin{eqnarray*}
 % \nonumber to remove numbering (before each equation)
   PQ &=& (P\eta_{x_1}, P\eta_{x_2}, \cdots, P\eta_{x_{r^m}}) \\
   &=& (\theta_{x_1}\eta_{x_1}, \theta_{x_2}\eta_{x_2}, \cdots, \theta_{x_{r^m}}\eta_{x_{r^m}}) \\
   &=& (\eta_{x_1}, \eta_{x_2}, \cdots, \eta_{x_{r^m}})\Lambda\\
   &=& Q\Lambda,
 \end{eqnarray*}
 where \begin{equation*}
 \Lambda=\left(
  \begin{array}{ccccc}
    \theta_{x_1} &   &  &  \\
    &  \theta_{x_2}&   &    \\
  &    & \ddots& \\
   & &    &  \theta_{x_{r^m}}
  \end{array}
\right) {\rm is~a~diagonal~matrix}.
\end{equation*}
Note that when $v=0$, \begin{equation*}
\theta_a=\begin{cases}
\emph{ }0,  &{\rm if}~a=0;\\
   \emph{ }\varphi(a^{-1})g(\varphi,\chi_1), &{\rm if}~a\in \mathbb{F}_{r^m}^*. \\
\end{cases}
\end{equation*}Let's just say $g:=g(\varphi,\chi_1)$ for convenience.

It is easy to know that \begin{equation}\label{eqd}
 \Lambda=g(\varphi,\chi_1)\left(
  \begin{array}{ccccccc}
    1 &   &  &&&&  \\
     & \ddots   &&&&& \\
     && 1  &&&&    \\
    &  &   &-1&&&    \\
  &    &&&\ddots&& \\
   & &    &&&-1&\\
   & &    &&&&0
  \end{array}
\right).
\end{equation}
%$\{\eta_a:=(\chi_a(x_1), \chi_a(x_2), \cdots, \chi_a(x_{r^m}))^T: a\in \mathbb{F}_{r^m}\}$.

%Let $C:=C_{(\varphi,v)}$ be a linear code over $\mathbb{F}_q$ with generator matrix $G=[I_{r^m}, P]$, where $P$ is defined by Section 3. Let $\mathbb{F}_{r^m}=\{x_i: 1\leq i\leq r^m\}$. Define the $r^m\times r^m$ matrix $P=(p_{ij})\in M_{r^m}(\mathbb{F}_q)$ by $p_{ij}=\rho(x_j-x_i)$. Then
%\begin{equation*}
%{P}=\left(
 % \begin{array}{ccccc}
  %  v &  \varphi(x_2-x_1) &\varphi(x_3-x_1)&  \cdots&  \varphi(x_{r^m}-x_1) \\
   %\varphi(x_1-x_2) &  v&   \varphi(x_3-x_2)& \cdots&  \varphi(x_{r^m}-x_2)   \\
   %\vdots &  \vdots & \vdots  & \ddots& \vdots \\
   %\varphi(x_1-x_{r^m})& \varphi(x_2-x_{r^m})&   \varphi(x_3-x_{r^m}) & \cdots &  v
  %\end{array}
%\right).
%\end{equation*}
%Hence, $G$ is a $r^m\times 2r^m$ matrix over $\mathbb{F}_q$, rank$(G)=r^m$ and
It follows the definition of the linear code $C$ that $C$ can be expressed in the following form:
\begin{eqnarray*}
% \nonumber to remove numbering (before each equation)
  C &=& \{c{(\bm{k})}=\bm{k}G, \bm{k}\in \mathbb{F}_q^{r^m}\}, {\rm where}~\bm{k}=(k_1, k_2, \cdots, k_{r^m}).
\end{eqnarray*}
For any codeword $c(\bm{k})$ in $C$, we have
\begin{eqnarray*}
% \nonumber to remove numbering (before each equation)
  c(\bm{k}) &=& \bm{k}G \\
  &=& \bm{k}(I_{r^m}, P)\\
   &=&(\bm{k},\bm{k}P)\\
    &=&(\bm{k},\bm{l}), {\rm where}~\bm{l}:=\bm{l}(\bm{k})=\bm{k}P\\
    &=&(k_1, k_2, \cdots, k_{r^m},l_1, l_2, \cdots, l_{r^m}).
\end{eqnarray*}
Multiply both sides of the equation $\bm{l}=\bm{k}P$ by the matrix $Q$, we obtain
$$\bm{l}Q=\bm{k}PQ=\bm{k}Q\Lambda.$$
Based on the above discussion and combined with Eq. (\ref{eqd}), we have the following three equations:
\begin{eqnarray}\label{eqcc}
% \nonumber to remove numbering (before each equation)
  (l_1-gk_1,\cdots, l_{r^m}-gk_{r^m})(\eta_{x_1},\cdots, \eta_{x_\frac{{r^m}-1}{2}}) &=&{\bm 0};
\end{eqnarray}
\begin{eqnarray}\label{eqc1c}
% \nonumber to remove numbering (before each equation)
 (l_1+gk_1,\cdots, l_{r^m}+gk_{r^m})(\eta_{x_\frac{{r^m}+1}{2}},\cdots, \eta_{x_{r^m-1}}) &=&{\bm 0};\\
 l_1+\cdots+l_{r^m}&=& {0}\nonumber.
\end{eqnarray}
%\begin{eqnarray}\label{eqcc}
% \nonumber to remove numbering (before each equation)
  %(l_1-gk_1,\cdots, l_{r^m}-gk_{r^m})(\eta_{x_1},\cdots, \eta_{x_\frac{{r^m}-1}{2}}) &=& {\bm 0}; \nonumber\\
 %  (l_1+gk_1,\cdots, l_{r^m}+gk_{r^m})(\eta_{x_\frac{{r^m}+1}{2}},\cdots, \eta_{x_{r^m-1}})&=& {\bm 0}; \\
  % l_1+\cdots+l_{r^m}&=& {0}\nonumber.
%\end{eqnarray}
In view of the above three equations, we present the following theorem. Before we do that, let's give some definitions. We define $\left(
  \begin{array}{c}
    \mu_{x_1} \\
  \vdots  \\
   \mu_{x_{r^m}}
  \end{array}
\right):=(\eta_{x_1}, \cdots, \eta_{x_\frac{{r^m}-1}{2}})$ and $\left(
  \begin{array}{c}
    \nu_{x_1} \\
  \vdots  \\
   \nu_{x_{r^m}}
  \end{array}
\right):=(\eta_{x_\frac{{r^m}+1}{2}}, \cdots, \eta_{x_{r^m-1}})$.
\begin{thm}\label{6.1} Let $C:=C_{(\varphi,0)}$ be a linear code over $\mathbb{F}_q$ with generator matrix $G=[I_{r^m}, P]$ defined by Section 3. Let $A$ be a positive integer. Assume that any $A$ vectors in $\{\mu_{x_1},\cdots,\mu_{x_{r^m}}\}$ are linearly independent and any $A$ vectors in $\{\nu_{x_1},\cdots,\nu_{x_{r^m}}\}$ are also linearly independent. Then $d_{min}(C)\geq A+1.$
\end{thm}
\begin{proof}Suppose that $c(\bm{k})$ is any codeword in $C$ which satisfies that wt$(c(\bm{k}))<A+1$. Note that $c(\bm{k})=(\bm{k},\bm{k}P)=(\bm{k},\bm{l})=(k_1,\cdots,k_{r^m},l_1,\cdots,l_{r^{m}})$. Set $\Omega:=\{(l_1,k_1),\cdots,(l_{r^m},k_{r^m})\}.$ Let $x=\#\{(l_i,k_i)\in \Omega\mid(l_i,k_i)=(0,0)\}$, $y=\#\{(l_i,k_i)\in \Omega\mid~{\rm Only~one~of}~l_i~{\rm and}~{k_i}~{\rm is}~0\}$ and $z=\#\{(l_i,k_i)\in \Omega\mid l_i\neq0~{\rm and}~k_i\neq0\}$. Then we have
\begin{equation}\label{eqe}
\begin{cases}
\emph{ }x+y+z=r^m\\
   \emph{ }2x+y>2r^m-A-1. \\
\end{cases}
\end{equation}
From (\ref{eqe}), we obtain
\begin{eqnarray}\label{eqf}
% \nonumber to remove numbering (before each equation)
  x &>& r^m-A-1.
\end{eqnarray}
%$x>t-A-1.$

Let $u_i=l_i-gk_i$ and $w_{i}=l_{i}+gk_{i}$, where $1\leq i\leq r^m.$ It follows Eqs. (\ref{eqcc}) and (\ref{eqc1c}) that
\begin{equation}\label{eqg}(u_1,\cdots,u_{r^m})\left(
  \begin{array}{c}
    \mu_{x_1} \\
  \vdots  \\
   \mu_{x_{r^m}}
  \end{array}
\right)=u_1\mu_{x_1}+\cdots+u_{r^m}\mu_{x_{r^m}}={\bm 0}\end{equation} and \begin{equation}\label{eqh}(w_1,\cdots,w_{r^m})\left(
  \begin{array}{c}
    \nu_{x_1} \\
  \vdots  \\
   \nu_{x_{r^m}}
  \end{array}
\right)=w_1\nu_{x_1}+\cdots+w_{r^m}\nu_{x_{r^m}}={\bm 0}.\end{equation}
According to (\ref{eqf}), it is easy to know that there are at least $r^m-A$ zeros in $u_1,\cdots, u_{r^m}$. Similarly, there are also at least $r^m-A$ zeros in $w_1,\cdots, w_{r^m}$. Without loss of generality, let's assume that $u_{A+1}=\cdots=u_{r^m}=0$. Combined Eq. (\ref{eqg}) and $\mu_{x_1},\cdots, \mu_{x_A}$ are linearly independent, then we have $u_1=\cdots=u_A=0$. Hence, we obtain $u_1=\cdots=u_{r^m}=0$. Similarly, we also deduce $w_1=\cdots=w_{r^m}=0$. Therefore, we get $l_1=\cdots=l_{r^m}=0$ and $k_1=\cdots=k_{r^m}=0$.
Then $c(\bm{k})$ is a zero codeword. That is to say, for any nonzero codeword ${c}$ in $C$, we have wt$({c})\geq A+1.$ So $d_{min}(C)\geq A+1.$ This completes the proof.
%From what has been discussed above, $c(\bm{k})=({\bm k}, {\bm l})={\bm 0}.$ Since $c(\bm{k})$ is any nonzero codeword, $d_{min}(C)\geq A+1.$ This completes the proof.
\end{proof}
\begin{rem}According to Theorem \ref{6.1}, we expect to find the largest $A$ that satisfies the assumption of Theorem \ref{6.1}. It is trivial that $A=1$ satisfies the assumption of Theorem \ref{6.1}. When $\frac{r^m-1}{2}\geq 2$ and $m=1$, it is easy to prove that $A=2$ satisfies the assumption of Theorem \ref{6.1}. Based on the a lot of examples we have tried by Magma, we guess that $A=\frac{r^m-1}{2}$ satisfies the assumption of Theorem \ref{6.1}. If this conjecture is correct, then $d_{min}(C)\geq \frac{r^m+1}{2}.$ But we fail to prove it. Thus we would like to put it here as an open problem.
%\begin{enumerate}
 % \item [(1)]
  %\item [(2)]
  %\item [(3)] When $\frac{r^m-1}{2}\geq 3$ or $m\neq 1$, it's an open problem what the maximum $A$ can take on satisfies the assumption of Theorem \ref{6.1}.
%\end{enumerate}
\end{rem}

%In a similar way, we continue to consider the subcode $C_i~(i\geq 5)$
%$wt(k_1,k_2, \cdots, k_{r^m})\geq5$
%and calculate the minimum distance of $C_i$. Based on the above discussion, we find that as $wt(k_1,k_2, \cdots, k_{r^m})$ gradually increases, the minimum weight of ${C_i}$ also gradually increases.
%Then we obtain the following corollary, which determines the minimum distance $d_{\rm min}(C)$ of the linear code $C$.

\begin{Conjecture}\label{Conj} Let $p>3$, $r$ be two distinct prime numbers and $m$ a positive integer. Assume that $N=2$. Let $C:=C_{(\varphi,v)}$ be a linear code over $\mathbb{F}_q$ with generator matrix $G=[I_{r^m}, P]$, where $P$ is defined by Section 3. Then
\begin{itemize}
  \item [(1)] When $v=0$, we have
  \begin{equation*}
d_{\rm min}(C)=\begin{cases}
\emph{ }3, ~~~~~~&{\rm if}~r^m=3;\\
   \emph{ }\frac{r^m+5}{2}, ~&{\rm if}~r^m\neq 3.\\
\end{cases}
\end{equation*}
  %$$d_{\rm min}(C)=\frac{r^m+5}{2}.$$
  \item [(2)] When $v\neq 0$ and $r^m\equiv1~({\rm mod}~4)$, we have
\begin{equation*}
d_{\rm min}(C)=\begin{cases}
\emph{ }\frac{r^m+1}{2}, ~&{\rm if}~v=\pm1;\\
   \emph{ }\frac{r^m+5}{2}, ~&{\rm if}~v\neq \pm1.\\
\end{cases}
\end{equation*}
\end{itemize}
\end{Conjecture}
\begin{rem} Example \ref{ex2} and some examples in Table \ref{table} can illustrate the validity of the above results. In fact, we have tried a lot of examples by Magma, the conjecture is also correct. But we fail to prove
it. Thus we would like to put it here as a conjecture.
%(2) By discussing in the same method as above, we didn't find the rule to follow in the case $r^m\equiv 1~({\rm mod}~4)$ and $v\neq 0$, so we fail to discuss the minimum distance of the linear code $C$ in this case. Interested readers can explore it.
\end{rem}
\section{Conclusion}
In this paper, we propose a general method to construct LCD codes and linear codes with one-dimensional hull, by using an analogue of Gauss sums where both the corresponding additive and multiplicative character take their
values in a finite field instead of the complex numbers. Based
on the eigenvalues of the matrix $PP^T$, some sufficient conditions for a linear code to be an LCD code (resp. a linear code with one-dimensional hull) have been presented
in this paper. With these conditions, we obtain some optimal and almost optimal LCD codes (resp. linear codes with one-dimensional hull) by Magma \cite{B}, which are exhibited
in Table \ref{table}. Additionally, we also obtain several almost MDS LCD codes (resp. almost MDS codes with one-dimensional hull) (see Examples \ref{exa3} and \ref{exa2}).

Compared with \cite{LZ}, their construction methods are specific and special, while our methods are more general and direct. It is mainly reflected in three aspects:

\begin{enumerate}
 \item In \cite{LZ}, the matrix $P$ studied by the authors satisfies the symmetry property, while the matrix $P$ we employed in this paper
is a general matrix whose eigenvalues are completely determined;
 \item In \cite{LZ}, the authors constructed linear codes with one-dimensional hull over finite fields by using the generator matrix over quadratic number fields,
while we construct them directly by utilizing the generator matrix over finite fields;
 \item Taking $N=2$, we obtain that \cite[Theorem 5]{LZ} is a special of our results in Theorem \ref{thm1} by comparing the constraints. The results of Theorem \ref{thm2}
contain \cite[Theorems 3 and 4]{LZ}. In some sense, some of linear codes with one-dimensional hull we constructed may be new when $N>2$ by comparing with \cite{LZ}
(see Example \ref{exa}).
In addition, we present a lower bound on the minimum distance of linear code $C$ over $\mathbb{F}_q$ with generator matrix $G=[I_{r^m}, P]$ when $N=2$.

\end{enumerate}

We should emphasize that our results apply to $(p,r)=1$. It would be interesting to extend the results of the present work to $p=r$. The main open problem is Conjecture \ref{Conj}. In addition, although there are many LCD codes and linear codes with one-dimensional hull, it seems to be difficult to determine the minimum distances of the codes presented in this paper when $N>2$. It will be of interest to find other constructions such that the minimum distances of these codes can be determined.


\begin{thebibliography}{1}
\bibitem{AK} E. F. Assmus, J. D. Key, Affine and projective planes, Discret. Math., vol. 83, nos. 2-3, pp. 161-187, 1990.
\bibitem{B} W. Bosma, J. J. Cannon, C. Fieker and A. Steel, Hand-book of Magma functions, Edition 2.22 5669 pages (2016).
http://magma.maths.usyd.edu.au/magma/.
\bibitem{CG} C. Carlet, S. Guilley, Complementary dual codes for counter-measures to side-channel attacks. In: E.R. Pinto et al. (eds.), Coding Theory and Applications. CIM Series in Mathematical Sciences, vol. 3, pp. 97-105, Springer (2014). J. Adv. Math. Commun., vol. 10, no. 1, pp. 131-150, 2016.
\bibitem{CLM} C. Carlet, C. J. Li, S. Mesnager, Linear codes with small hulls in semi-primitive case, Des. Codes Cryptogr., vol. 87, pp. 3063-3075, 2019.
\bibitem{CMTQ} C. Carlet, S. Mesnager, C. M. Tang, Y. F. Qi, R. Pellikaan, Linear codes over $\F_q$ are equivalent to LCD codes for $q>3$. IEEE Trans. Inf. Theory, vol. 64, pp. 3010-3017, 2018.
\bibitem{CMTQ1} C. Carlet, S. Mesnager, C. M. Tang, Y. F. Qi, Euclidean and Hermitian LCD MDS codes. Des. Codes Cryptogr., vol. 86, pp. 2605-2618, 2018.
\bibitem{CMTQ2} C. Carlet, S. Mesnager, C. M. Tang, Y. F. Qi, New characterization and parametrization of LCD codes. IEEE Trans. Inf. Theory, vol. 65, 39-49, 2019.
\bibitem{G} M. Grassl, Bounds on the minimum distance of linear codes and quantum codes, http://www.codetables. de (2019). Accessed 2 Jan 2019.
%\bibitem{GJG} K. Guenda, S. Jitman, T. Gulliver, A: Constructions of good entanglement-assisted quantum error correcting codes. Des. Codes Cryptogr., vol. 86, 121-136, 2018.
\bibitem{J} L. F. Jin, Construction of MDS codes with complementary duals. IEEE Trans. Inf. Theory, vol. 63, pp. 2843-2847, 2017.
\bibitem{J1} L. F. Jin, C. P. Xing, Algebraic geometry codes with complementary duals exceed the asymptotic Gilbert-Varshamov bound, IEEE Trans. Inform. Theory, vol. 64, pp. 6277-6282, 2018.
\bibitem{JL} S. Jitman, and S. Ling, Quasi-abelian codes, Des. Codes Cryptogr., {\bf 74}, (2015),  511--531.
\bibitem{L1} J. S. Leon, Computing automorphism groups of error-correcting codes, IEEE Trans. Inform. Theory, vol. 28, pp. 496-511, 1982.
\bibitem{L2} J. S. Leon, Permutation group algorithms based on partition, I: theory and algorithms, J. Symbolic Comput., vol. 12, pp. 533-583, 1991.
\bibitem{LL} X. S. Liu, H. L. Liu, LCD codes over finite chian rings, Finite Fields Their Appl., vol.
34, pp. 1-19, 2015.
\bibitem{LL1} X. S. Liu, Y. Fan, H.L. Liu, Galois LCD codes over finite fields. Finite Fields Their
Appl., vol. 49, pp. 227-242, 2018.
%\bibitem{LC} G. J. Luo, X. W. Cao, X. J. Chen, MDS codes with hulls of arbitrary dimensions and their quantum error correction. IEEE Trans. Inf. Theory, vol. 65, 2944-2952, 2019.
\bibitem{ST} S. Mesnager, C. M. Tang, Y. F. Qi, Complementary dual algebraic geometry codes, IEEE Trans. Inf. Theory, vol. 64, no. 4, pp. 2390-2397, Apr. 2018.

\bibitem{LNC} R. Lidl, H. Niederreiter, P. M. Cohn, \emph{Finite fields}, Cambridge University Press, 1997.
\bibitem{LZ} C. J. Li, P. Zeng, Constrctions of linear codes with one-dimensional hull, IEEE Trans. Inf. Theory, vol. 65, no. 3, pp. 1668-1676, 2019.
\bibitem{M} J. L. Massey, Linear codes with complementary duals, Discrete Math., vols. 106-107, pp. 337-342, Sep. 1992.
\bibitem{QC1} L. Q. Qian, X. W. Cao, S. Mesnager, Linear codes with one-dimensional hull associated with Gaussian sums, Cryptogr. Commun. (2020). https://doi.org/10.1007/s12095-020-00462-y.
\bibitem{QC2} L. Q. Qian, X. W. Cao, Linear complementary dual codes constructed by general Gaussian sums over finite fields, submitted.
\bibitem{S1} N. Sendrier, Finding the permutation between equivalent codes: the support splitting algorithm, IEEE Trans. Inform. Theory, vol. 46, pp. 1193-1203, 2000.
\bibitem{SH} M. J. Shi, D. T. Huang, L. Sok, P. Sol\'e, Double circulant LCD codes over $\mathbb{Z}_4$, Finite Fields Their Appl., vol. 58, pp. 133-144, 2019.
\bibitem{LS} L. Sok, M. J. Shi, P. Sol\'e, Constructions of optimal LCD codes over large finite fields, Finite Fields Their Appl., vol. 50, pp. 138-153.
\bibitem{S2} N. Sendrier, G. Skersys, On the computation of the automorphism group of a linear code, in: Proceedings of IEEE ISIT¡¯2001, Washington, DC, 2001, p. 13.


\end{thebibliography}
\end{document}